\documentclass{article}
\usepackage{spconf,amsmath,amssymb,graphicx,booktabs,xcolor, hyperref}
\usepackage{multirow}
\usepackage{makecell}
\usepackage{diagbox}
\usepackage{amsthm}
\theoremstyle{definition}
\newtheorem{definition}{Definition}

\newcommand{\AS}{\mathcal{AS}}   

\newcommand{\eps}{\epsilon}
\newcommand{\FPR}{\mathrm{FPR}}
\newcommand{\TPR}{\mathrm{TPR}}
\title{VoxGuard: Evaluating user and attribute privacy in speech via Membership Inference Attacks}
\name{
    \begin{tabular}{@{}c@{}}
    Efthymios Tsaprazlis, Thanathai Lertpetchpun, Tiantian Feng, \\ \textit{Sai Praneeth Karimireddy}, \textit{Shrikanth Narayanan}
    \end{tabular}
    }
\address{University of Southern California, Los Angeles, CA}
\usepackage{fancyhdr}
\fancypagestyle{firstpage}{
\fancyhf{}
\fancyfoot[C]{\footnotesize Copyright 2026 IEEE. Published in ICASSP 2026 - 2026 IEEE International Conference on Acoustics, Speech and Signal Processing (ICASSP), scheduled for 3-8 May 2026 in Barcelona, Spain. Personal use of this material is permitted. However, permission to reprint/republish this material for advertising or promotional purposes or for creating new collective works for resale or redistribution to servers or lists, or to reuse any copyrighted component of this work in other works, must be obtained from the IEEE. Contact: Manager, Copyrights and Permissions / IEEE Service Center / 445 Hoes Lane / P.O. Box 1331 / Piscataway, NJ 08855-1331, USA. Telephone: + Intl. 908-562-3966.}

}
\begin{document}
\ninept
\maketitle
\begin{abstract}
Voice anonymization aims to conceal speaker identity and attributes while preserving intelligibility, but current evaluations rely almost exclusively on Equal Error Rate (EER) that obscures whether adversaries can mount high-precision attacks. We argue that privacy should instead be evaluated in the low false-positive rate (FPR) regime, where even a small number of successful identifications constitutes a meaningful breach. To this end, we introduce \textbf{VoxGuard}, a framework grounded in differential privacy and membership inference that formalizes two complementary notions: \emph{User Privacy}, preventing speaker re-identification, and \emph{Attribute Privacy}, protecting sensitive traits such as gender and accent. Across synthetic and real datasets, we find that informed adversaries, especially those using fine-tuned models and max-similarity scoring, achieve orders-of-magnitude stronger attacks at low-FPR despite similar EER. For attributes, we show that simple transparent attacks recover gender and accent with near-perfect accuracy even after anonymization. Our results demonstrate that EER substantially underestimates leakage, highlighting the need for low-FPR evaluation, and recommend \texttt{VoxGuard} as a benchmark for evaluating privacy leakage.
\end{abstract}
\vspace{-1mm}
\section{Introduction}
\vspace{-1mm}

Speech recognition systems raise significant privacy concerns because spoken data reveals both linguistic content and biometric identity. Regulatory frameworks, such as GDPR \cite{gdpr2016}, explicitly classify voice as personal data and, in some cases, as biometric data. This dual nature of voice raises an important question: what types of sensitive information can be inferred from speech? On the one hand, the linguistic channel conveys the literal content of what is said. On the other hand, the paralinguistic channel, such as tone, pitch, and timbre, forms a distinctive "voiceprint" that can uniquely identify a speaker~\cite{zhang2023voicepm}. Apart from speaker identification, voice also encodes a wide range of private attributes such as gender, accents, emotional state, and even health conditions~\cite{haq2009speaker,tayebi2024addressing, ahangaran2025obfuscation}, which individuals may prefer to keep confidential. In this work, we direct part of our analysis toward these personal attributes.

In recent years, the speech processing community has advanced \emph{voice anonymization} methods and challenges, commonly evaluating speaker privacy through the degradation of Automatic Speaker Verification (ASV) performance, most often via the Equal Error Rate (EER) metric \cite{tomashenko2024voiceprivacy}. However, EER captures only an \emph{average-case} operating point and can obscure individual risks and tail behavior. In fact, EER primarily reflects the adversary’s weakest decisions rather than their strongest \cite{nautsch2020privacy}. Privacy, by its nature, cannot be fairly evaluated through average-case measures. We argue that more robust assessments should quantify \emph{worst-case leakage}, particularly in the \emph{low false positive rate (FPR)} regime, where an adversary can successfully identify individuals without plausible deniability \cite{carlini2022membership}.

This paper uses viewpoints from Differential Privacy, particularly on membership inference, to evaluate speech privacy. We introduce \textbf{\texttt{VoxGuard}}, an evaluation framework consisting of two complementary notions: \textit{User Privacy} corresponding to the traditional notion of preventing speaker re-identification and \textit{Attribute Privacy} quantifying the risk of inferring speaker traits (e.g., gender, accent). For each, we define strict- and relaxed-case notions aligned with DP-style indistinguishability and evaluate at low-FPR operating regions to expose worst-case leakage. Empirically, we find that EER can diverge from measuring privacy at low-FPR regimes and we provide upper-bounds for the attacks.

Our contributions: 
i) We reframe speech privacy as a membership inference problem and formalize two complementary definitions inspired by Differential Privacy: \emph{User Privacy} and \emph{Attribute Privacy}. ii) We argue that privacy in speech should be evaluated in the \emph{low-FPR regime} and not by EER. Even a few reliable breaches can constitute a privacy violation. iii) Our results reveal that ASV models with comparable EERs can differ by orders of magnitude in the low-FPR regime, exposing much greater privacy risks than average-case metrics suggest. iv) Speaker attributes remain unprotected, as simple transparent attacks can fully expose traits such as gender and accent, underscoring the need for explicit defenses.



\vspace{-1mm}
\section{Related Works}
\vspace{-1mm}
\subsection{Membership inference and differential privacy}
Membership Inference Attacks (MIA) are a standard approach, in which an adversary tries to determine whether a specific entry is in a dataset, for quantifying privacy leakage 
in machine learning research \cite{shokri2017membership}. Carlini \textit{et al.}~\cite{carlini2022membership} emphasized that 
evaluating MIA at very low FPR provides a more faithful picture of worst-case 
privacy leakages than average-case metrics such as accuracy. This perspective aligns with the principles of DP~\cite{dwork2006differential}, which formally bound the influence of any single individual’s data on the output distribution of a mechanism. DP has recently 
been explored in speech applications, such as in speech emotion recognition using Federated Learning \cite{feng2022user} or embedding perturbation \cite{wang2024asynchronous,shamsabadi2022differentially}, but direct integration 
remains challenging due to utility trade-offs. Our framework instead uses DP analytically, and we adopt MIA evaluation in the low-FPR regime as a practical proxy for bounding worst-case leakage \cite{carlini2022membership}. This bridges the gap between the existing privacy metric in ASV and  DP.

\vspace{-1mm}
\subsection{Privacy metrics in speech}
In speech processing, privacy has been primarily evaluated through degradation of ASV performance. In contrast to prior metrics such as EER \cite{tomashenko2024voiceprivacy} or $C_{\ell\ell r}$ \cite{noe2021adversarial} that capture an average-case attacker performance, our evaluation is explicitly grounded in membership inference on low-FPR regimes, which better reflect worst-case leakage. Approaches like ZEBRA~\cite{nautsch2020privacy} and GDPR-inspired linkability/singling-out \cite{vauquier2025legally} also move toward worst-case leakage, but remain tied to ASV operating points. By framing both identity and attribute privacy in a DP-style $\eps$ formulation, our metrics directly quantify the maximum information gain available to an attacker, rather than the relative degradation of ASV systems. This allows us to capture leakage that EER or linkability may miss, while still being comparable with traditional ASV-based reporting.

\vspace{-1mm}
\section{Method}

In MIA, the adversary $\mathcal{A}$ aims to distinguish between two neighboring datasets $D$ and $D'$. Dataset $D$ defines an enrollment set $E$ based on the value of interest (e.g., a particular speaker identity or a specific attribute) to the attacker, while $D'$ corresponds to a modified enrollment set in which there is exactly one difference in that value of interest. The adversary receives the system output $\AS(\cdot)$ and must decide whether it originated from $D$ or from $D'$.

\subsection{User Privacy}
In this case, neighboring datasets differ in the enrolled \emph{user identity}. Formally, $D$ contains an enrollment set for user $i$, while $D'$ contains an enrollment set for a different user $j$ instead of the user $i$. Membership inference then asks: \emph{did the probe utterance come from the enrolled user or not?} This setting aligns directly with automatic speaker verification, but is reframed as a membership inference problem. Under \emph{User Privacy}, we evaluate privacy by examining how distinguishable the adversary’s outputs are when the enrolled identity changes, expressed through the DP style bounds.

\begin{definition}[\textbf{($\eps, \delta$)-User Privacy}]
A random mechanism $\AS$ satisfies ($\eps, \delta$)-User Privacy, where $\eps >0$, $\delta \in [0,1)$ iff for any distinct speakers $i\neq j$ and any text $t^{(i)}$, $t^{(j)}$,
\begin{equation}
\Pr\!\left[\AS(i, t^{(i)})=s\right] \;\le\; e^{\eps}\; \Pr\!\left[\AS(j, t^{(j)})=s\right] \;+\; \delta,\;\forall s.
\label{eq:user}
\end{equation}
\end{definition}
\vspace{-0.5mm}

\noindent For the \emph{strict case}, we assume that the text $t$ is identical for both speakers $i$ and $j$. Fixing $t$ isolates identity as the primary varying factor, yielding a strict indistinguishability requirement. For the \emph{relaxed case}, we let $t^{(i)} \sim \mathcal{D}_i$ and $t^{(j)} \sim \mathcal{D}_j$ denote texts drawn from user-specific distributions. This relaxes the requirement to a more realistic assumption where content naturally varies.




\subsection{Attribute Privacy}

Here, neighboring datasets differ in the value of a selected \emph{attribute} associated with the enrollment. For example, $D$ might encode an enrollment with attribute ``female,'' while $D'$ encodes ``male.'' Membership inference in this case asks: \emph{does the probe utterance exhibit the target attribute or not?} This captures leakage of sensitive characteristics such as gender, accent, or age, and generalizes initial approaches that explored speaker attribute inference in federated learning~\cite{feng2022user}. By framing attribute prediction as a membership game, \emph{Attribute Privacy} quantifies how much the output distribution shifts when the attribute value is flipped.

Let $\mathbf{a}_i$ denote the attribute vector of speaker $i$, consisting of values such as gender and accent. The length of $\mathbf{a}_i$ equals the number of attributes under consideration (two in our case). For the attribute at position $k$, we define a modified vector $\tilde{\mathbf{a}}_i$ as
\begin{equation}
\tilde{\mathbf{a}}_i[m] =
\begin{cases}
\mathbf{a}_i[m], & m\neq k,\\
u, & m=k,
\end{cases}
\end{equation}
for some alternate value $u$ for the attribute $\mathbf{a}_i[k]$.

\begin{definition}[\textbf{Strict ($\eps$,$\delta$)-Attribute Privacy}]
A random mechanism $\AS$ satisfies strict ($\eps, \delta$)-Attribute Privacy, where $\eps >0$, $\delta \in [0,1)$ iff for any $i$, any distinct $u \neq \mathbf{a}_i[k]$ and any texts $t$,
\begin{equation}
\Pr\!\left[\AS(\mathbf{a}_i, t)=s\right]\;\le\; e^{\eps}\;\Pr\!\left[\AS(\tilde{\mathbf{a}}_i, t)=s\right] \;+\; \delta,\;\forall s.
\label{eq:attr-worst}
\end{equation}
\end{definition}
\noindent This captures indistinguishability under a \emph{single-attribute} change, holding other attributes fixed.
However, in real-world settings, we argue that attributes can be entangled with each other, making it impossible to modify only one while keeping other attributes the same. For this reason, we relax the strict privacy assumption and move more to group comparison. 

\begin{definition}[\textbf{Relaxed ($\eps$,$\delta$)-Attribute Privacy}]
A random mechanism $\AS$ satisfies relaxed ($\eps, \delta$)-Attribute Privacy, where $\eps >0$, $\delta \in [0,1)$ iff for any $i$, any distinct $u \neq \mathbf{a}_i[k]$ and any texts $t$,
\begin{equation}
\Pr\!\left[\AS(\mathbf{a}_i[k], t)=s\right]\;\le\; e^{\eps}\;\Pr\!\left[\AS(u_i, t)=s\right] \;+\; \delta,\;\forall s.
\label{eq:attr-avg}
\end{equation}
\end{definition}

\begin{figure}
    \centering
    \includegraphics[width=0.9\linewidth]{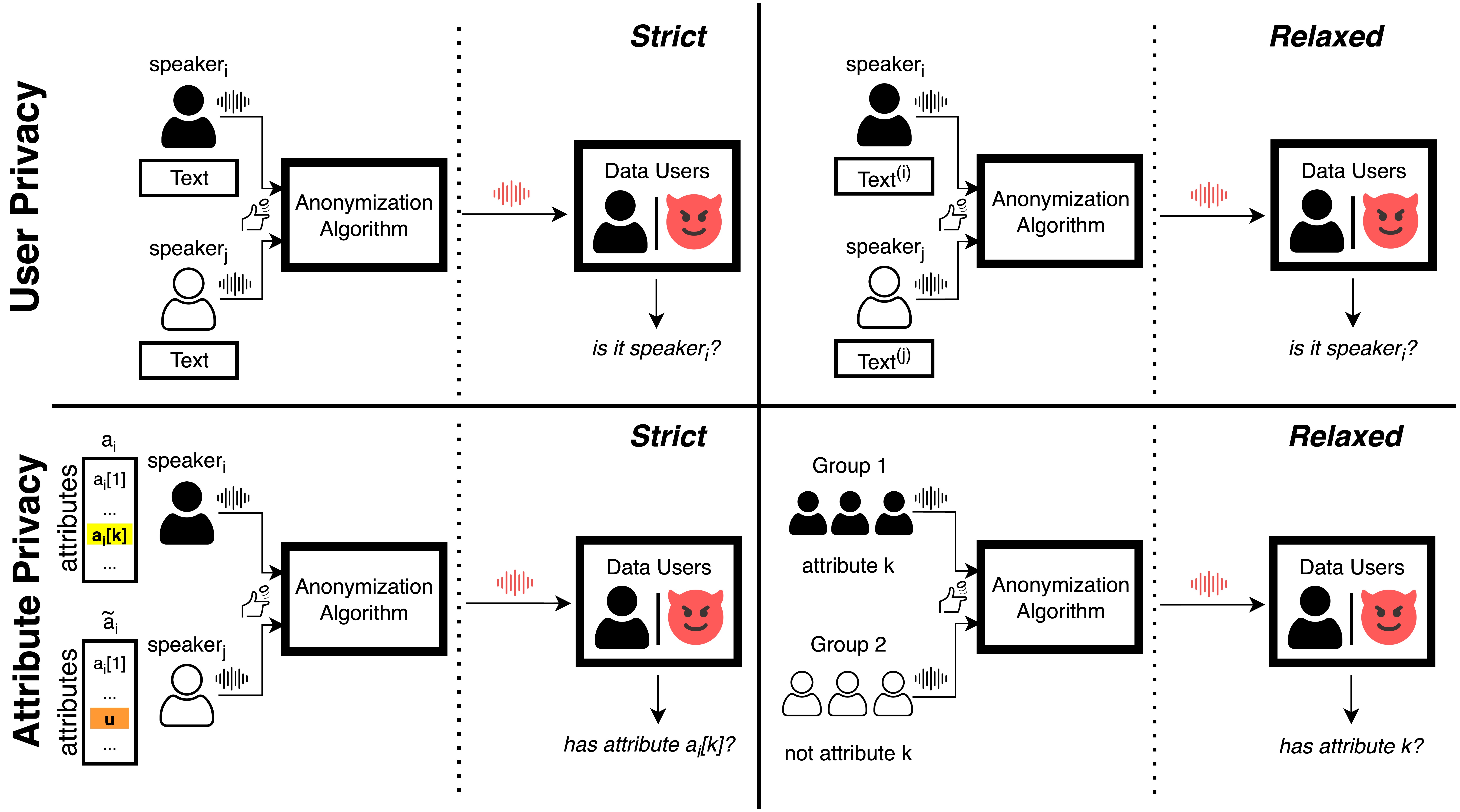}
    \caption{Illustration of the four \textsc{\texttt{VoxGuard}} evaluation tasks. 
\textbf{\textsc{User-Privacy-Strict}}: adversary compares same-text pairs to decide if two samples belong to the same speaker (identity as the varying factor). 
\textbf{\textsc{User-Privacy-Relaxed}}: adversary compares arbitrary texts from two speakers, reflecting the general case where content varies. 
\textbf{\textsc{Attr-Privacy-Strict}}: pairs are matched on all attributes except one, isolating the effect of a single attribute. 
\textbf{\textsc{Attr-Privacy-Relaxed}}: pairs differ in one attribute without controlling for others, modeling the general case where multiple traits vary.}
    \label{fig:tasks}
\vspace{-5mm}
\end{figure}

\noindent In both definitions, the DP inequalities provide a way to quantify indistinguishability. The smaller the gap between $\Pr[\mathcal{A}S(D)=s]$ and $\Pr[\mathcal{A}S(D')=s]$, the weaker the adversary’s ability to win the game, and thus the stronger the privacy guarantee. Importantly, while \textit{User Privacy} focuses on \emph{identity-level membership}, \textit{Attribute Privacy} focuses on \emph{feature-level membership},  allowing \texttt{VoxGuard} to capture complementary aspects of privacy leakage.
In Figure \ref{fig:tasks}, we present the \texttt{VoxGuard} tasks derived from the above definitions.



\subsection{Membership Inference and Low-FPR Metrics}
In all four settings, evaluation reduces to deciding whether two anonymized utterances have the \emph{same} identity or share the \emph{same} attribute value. Let a decision rule achieve $(\FPR,\TPR)$ on a balanced trial set. The worst-case privacy loss at that operating point is upper-bounded by the log-likelihood ratio \cite{thudi2022bounding}:
\vspace{-0.5mm}
\begin{equation}
\hat\eps\;=\;\max\!\left\{\ln\left(\frac{\TPR}{\FPR}\right),\ln\left(\frac{1-\FPR}{1-\TPR}\right)\right\}
\label{eq:eps-approx}
\end{equation}
\vspace{-0.5mm}
motivating reports of $\TPR$ at \emph{low} $\FPR\in\{10^{-2},10^{-3},10^{-4}\}$ and the minimal $\FPR$ achieving a non-trivial $\TPR$. While EER summarizes an average operating point, low FPR behavior better reflects worst-case leakage.


    
    
    

\vspace{-2mm}
\section{Evaluation}

\subsection{Experimental Setup}

\vspace{-1mm}
\subsubsection{Synthesized Speech}
We use the pre-trained XTTS \cite{casanova2024xtts} from the Coqui project~\footnote{\url{https://github.com/coqui-ai}} to synthesize speech and use the first 10k transcripts from \mbox{train-clean-100} of LibriTTS-R~\cite{koizumi2023libritts}. 
We control speaker attributes of synthesized speech by using the source utterances from the test-set of Vox-Profile \cite{feng2025vox}. In this paper, we focus on two speaker attributes: gender (male and female) and accent (North American, British, and South Asia), which are independent of each other. Six speakers are selected from the Vox-Profile test set and have the probability of Vox-Profile to that desired accent more than 95\% and low background noise. We synthesized a total of 60K utterances consisting of 10K utterances per speaker creating the \textit{Synthetic-X} dataset. We also use Kokoro TTS~\footnote{\url{https://github.com/hexgrad/kokoro}} to synthesize another synthesized dataset from the same transcript, conditioned on \texttt{af\_heart}, \texttt{am\_adam}, and \texttt{am\_echo} speaker embeddings. We refer to this dataset as \textit{Synthetic-K}.

\vspace{-1mm}
\subsubsection{Real Speech}
For evaluation on real speech data, we use the VCTK corpus \cite{yamagishi2019vctk} and ParaSpeechCaps \cite{diwan2025scaling}. For attribute privacy analysis, we unify the accent labels from both datasets into three coarse categories that align with the ``Narrow Accent'' definition in Vox-Profile. Specifically, we map \textit{North American}, \textit{American}, and \textit{Canadian} accents into the \textbf{North American} group; \textit{British}, \textit{Welsh}, and \textit{Scottish} into the \textbf{British} group; and \textit{Indian} and \textit{Indian-American} into the \textbf{South Asia} group. This grouping enables consistent attribute comparisons across the two datasets.

\vspace{-1mm}
\subsubsection{Model and Evaluation Details}
We adopt the VoicePrivacy 2024 \cite{tomashenko2024voiceprivacy} baseline B5 (ASR-BN)\cite{champion2023anonymizing} system for speech anonymization. 
This pipeline leverages both fundamental frequency (F0) and VQ bottleneck (VQ-BN) features from an ASR model 
trained to predict left-biphones. These features, combined with a designated speaker embedding, are passed to a HiFi-GAN vocoder to synthesize the anonymized waveform.  
All data are anonymized four times with different speakers of desired speaker attributes to generate multiple positive examples for the \textsc{Attr-Privacy-Strict} task, ensuring sufficient coverage of attribute-controlled scenarios. Throughout our evaluation, we assume the privacy attacker to be an \emph{Informed Attacker}. The adversary has full access to the anonymization system and can independently generate anonymized data.  We consider two attacker variants: one using a \textbf{Pretrained-Informed} embedding model (without adaptation) and one using a \textbf{Fine-tuned-Informed} model that is further adapted on anonymized data. Evaluating privacy metrics under these two attackers allows us to assess whether the effectiveness of privacy anonymization is overestimated when attacker adaptation is ignored.

For evaluation, we focus on the low-FPR regime, a standard practice in computer security \cite{pantel1998spamcop, kantchelian2015better,ho2017detecting,pantel1998spamcop}. The intuition is that even at very low false positive rates, an attacker who achieves non-trivial true positives can reliably compromise intended privacy anonymization for at least a small subset of users \cite{carlini2022membership}. In a sensitive domain such as speech, this already constitutes a successful privacy breach. Accordingly, in our experiments we compute the ROC curve and report three metrics: the EER, the TPR at low-FPR, and the worst-case privacy loss~$\eps$. We estimate $\eps$ using Eq.~\eqref{eq:eps-approx}, we fix $\delta=10^{-4}$ and confidence intervals are obtained via $200\times$ bootstrap resampling over half of the available trials.

\subsection{User Privacy Attacks}
For user privacy evaluation, we extract ECAPA-TDNN \cite{desplanques2020ecapa} speaker embeddings from anonymized utterances and compute cosine similarity between enrollment and test embeddings as the verification score. We consider two attacker variants:
  \textbf{Pretrained-Informed:} An off-the-shelf ECAPA-TDNN model without adaptation.  
 \textbf{Fine-tuned-Informed:} The same ECAPA-TDNN model fine-tuned on anonymized 
  data in a cross-domain setting (\textit{Synthetic-K} $\rightarrow$ \textit{Synthetic-X}; 
  VCTK \cite{yamagishi2019vctk} $\rightarrow$ ParaSpeechCaps~\cite{diwan2025scaling}), simulating an informed adversary adapted 
  to the anonymization process.  
We construct $100$k balanced trials per dataset,  scenario, and for enrollment lengths $L \in \{1,3,30,60,200\}$ per speaker, following \cite{vauquier2025legally} and extending beyond $L=30$. For $L>1$, we evaluate two scoring strategies: (i) \textbf{average-similarity}, where the enrollment vector is the mean of the $L$ segments (as in \cite{vauquier2025legally}), and (ii) \textbf{max-similarity}, where the score is the maximum cosine similarity between the probe and any enrollment embedding. The max-similarity evaluation better models a worst-case attacker, since averaging can dilute strong matches while max-similarity preserves them.

\subsection{Attribute Privacy Attacks}
For attribute privacy, we experiment with two speaker attributes: gender and accent. Attribute embeddings are extracted from anonymized speech using Vox-Profile models \cite{feng2025vox} (Whisper-based narrow-accent and age/sex). We then consider two attacker variants:
  \textbf{Cosine-similarity attacker} uses cosine similarity between enrollment and test attribute embeddings to decide whether two utterances share the same attribute. This simple baseline produced near-random separability in our experiments for both pretrained and finetuned embedding models.  
  \textbf{Transparent learned attacker:} trains logistic regression classifiers on 
  \emph{pairwise features} (embedding differences or concatenations\cite{feng2022user}) to classify same vs.\ 
  different attributes. We evaluate both \emph{in-distribution} (train/test split within the 
  same dataset) and \emph{cross-dataset} (train on anonymized VCTK, test on anonymized 
  ParaSpeechCaps) scenarios. We also explore fine-tuning the attribute encoders on anonymized 
  data.  
For each attribute and scenario (worst- and average-case), we construct $20$k balanced trials, and for enrollment we using lengths $L \in \{1,3,30,60,200\}$.  

\begin{table*}[t]
\footnotesize
\centering
\caption{Evaluation of user privacy attacks across conversation lengths $L$ and 
enrollment strategies (Avg vs.\ Max). We report Equal Error Rate (EER, ↓), 
worst-case $\hat{\epsilon}$ (↑), and True Positive Rate at 0.1\% and 0.01\% 
FPR (↑). For synthetic data, we report only $L \in \{1,3\}$ since each speaker has only four anonymized same-text utterances, allowing at most three for enrollment and one as probe.}
\vspace{0.5em}
\resizebox{\linewidth}{!}{
\begin{tabular}{llccccc}
\toprule
\textbf{Task} & \textbf{Strategy} & \textbf{$L=1$} & \textbf{$L=3$} & 
\textbf{$L=30$} & \textbf{$L=60$} & \textbf{$L=200$} \\
\midrule
\multicolumn{7}{c}{\textit{Each cell reports: EER (\%)$\;\mid\;$ $\hat{\epsilon}$ $\;\mid\;$ TPR@0.1\%FPR (\%)$\;\mid\;$ TPR@0.01\%FPR (\%)}} \\
\midrule
  & PT–Avg   & \multirow{2}{*}{42.46 $\mid$ 3.63 $\mid$ 0.96 $\mid$ 0.46}  & 38.99 $\mid$ 4.49 $\mid$ 1.44 $\mid$ 0.53 & \multicolumn{3}{c}{\multirow{4}{*}{\diagbox[width=40em,height=4.5em]{}{}}}\\
  \textsc{User-Privacy-Strict} & PT–Max   & & 38.68 $\mid$ 5.08 $\mid$ 1.59 $\mid$ 1.32  & & &   \\
  (Synthetic-X) & FT–Avg    & \multirow{2}{*}{39.95$\mid$ 3.35 $\mid$ 8.72 $\mid$ 0.48} & 37.27 $\mid$ 4.01 $\mid$ 0.57 $\mid$ 0.18 &  & &  \\
  & FT–Max    & & 37.10 $\mid$ 4.57 $\mid$ 1.80 $\mid$ 1.50 & & & \\
\midrule
  & PT–Avg   & \multirow{2}{*}{49.44 $\mid$ 0.74 $\mid$ 0.20 $\mid$ 0.00} & 49.10 $\mid$ 0.38 $\mid$ 0.09 $\mid$ 0.00 & 47.16 $\mid$ 0.67 $\mid$ 0.20 $\mid$ 0.00 & 46.47 $\mid$ 1.02 $\mid$ 0.20 $\mid$ 0.00 & 47.41 $\mid$ 1.08 $\mid$ 0.31 $\mid$ 0.00\\
  \textsc{User-Privacy-Relaxed} & PT–Max   &  & 49.50 $\mid$ 1.05 $\mid$ 0.17 $\mid$ 0.08 & 48.46 $\mid$ 1.28 $\mid$ 0.27 $\mid$ 0.08  & 47.73 $\mid$ 2.32 $\mid$ 0.41 $\mid$ 0.19 & 47.98 $\mid$ 3.15 $\mid$ 0.78 $\mid$ 0.49 \\
 (ParaSpeechCaps) & FT–Avg    & \multirow{2}{*}{43.11 $\mid$ 4.21 $\mid$ 1.66 $\mid$ 1.08} & 40.01 $\mid$ 3.88 $\mid$ 1.93 $\mid$ 0.76 & 36.30 $\mid$ 3.17 $\mid$ 1.15 $\mid$ 0.18 & 35.93 $\mid$ 3.15 $\mid$ 1.17 $\mid$ 0.40 & 36.53 $\mid$ 3.16 $\mid$ 1.16 $\mid$ 0.28 \\
  & FT–Max    &  & 40.89 $\mid$ 5.18 $\mid$ 3.40 $\mid$ 2.87 & 35.09 $\mid$ 7.25 $\mid$ 19.31 $\mid$ 18.83 & 31.29 $\mid$ 7.78 $\mid$ 32.29 $\mid$ 27.97 & 20.75 $\mid$ 8.40 $\mid$ 58.72  $\mid$ 52.82 \\
\bottomrule
\end{tabular}
}
\vspace{-4mm}
\label{tab:user_eval}
\end{table*}

\vspace{-1mm}
\section{Results}

\vspace{-1mm}
\subsection{User Privacy}

In Table~\ref{tab:user_eval}, we report performance across user-privacy attacks. While EER values fall within a relatively narrow range, examining the low-FPR regime reveals severe differences in attacker strength.
First, we note that EER alone can be misleading as a standalone privacy metric. For example, in \textsc{User-Privacy-Relaxed}, FT–Avg (Fine-tuned-Informed, Average Similarity) at $L=60$ achieves an EER of 36.3\%, suggesting relatively weak attacker performance. Yet, FT–Max (Fine-tuned-Informed, Max Similarity) at the same $L$ obtains a comparable EER (35.1\%) but also reaches TPR = 18.8\% at $\text{FPR}=10^{-4}$, over two orders of magnitude stronger than FT–Avg ($\approx$ 0.18\%). This demonstrates that comparable EER values can mask drastically different risks in practice, with low-FPR analysis exposing high-confidence membership inference attacks that EER fails to capture.

Second, the enrollment strategy matters. Max similarity consistently outperforms Avg similarity, particularly at longer L. For instance, under \textsc{User-Privacy-Relaxed} with FT–Max at $L=200$, the attacker achieves TPR = 58.7\% at $\text{FPR}=10^{-4}$, compared to only 1.2\% for FT–Avg. This reflects that averaging embeddings dilutes strong matches, whereas Max scoring preserves them.

\begin{figure}
    \centering
    \includegraphics[width=0.75\linewidth]{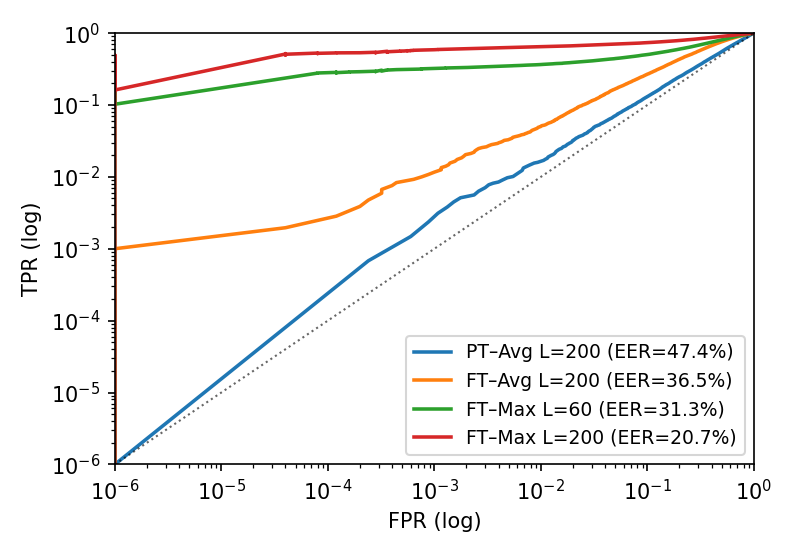}
    \caption{Comparing ROC curves in log–log scale reveals severe gaps between \textit{User Privacy} attacks. While average-case EERs range only from 20–47\%, their performance at low FPR differs drastically. FT–Max at $L=200$ achieves TPR $>$50\% at $\text{FPR}=10^{-4}$, whereas PT–Avg remains near random. This shows that reporting only EER obscures strong privacy leaks. Max scoring and fine-tuned attackers yield orders-of-magnitude stronger attacks in the low-FPR regime.}
    \label{fig:roc_loglog}
    \vspace{-4mm}
\end{figure}

Third, attacker adaptation amplifies leakage. Fine-tuned (FT) attackers consistently outperform pretrained (PT) attacking baselines. Even with short enrollments ($L=3$), FT–Max achieves TPR $\approx$ 3.4\% at $\text{FPR}=10^{-3}$, while PT–Avg remains nearly random. At larger $L$, fine-tuning compounds with Max scoring to expose high-confidence leakage, further underscoring the risks of ignoring attacker adaptation. These trends are clearly illustrated in Figure~\ref{fig:roc_loglog}.

Finally, results from \textsc{User-Privacy-Strict} confirm that controlling for lexical variability strengthens privacy attacks. Even in this constrained setting, FT attackers outperform PT by a wide margin, and Max scoring again amplifies leakage beyond what EER would suggest. This indicates that privacy risks persist both under realistic conditions with varying text (\textsc{User-Privacy-Relaxed}) and in controlled settings where only identity varies (\textsc{User-Privacy-Strict}).

\subsection{Attribute Privacy}

For attribute privacy, we begin with the same evaluation protocol as in user-privacy tasks: extracting embeddings and computing cosine similarity between enrollment and probe vectors. However, this baseline produced largely random results, with attackers often performing worse than chance. Even after fine-tuning the embedding models on anonymized VCTK data, performance remained unstable, suggesting that direct cosine-similarity is insufficient for reliable attribute inference under anonymization. Therefore, we shift to transparent attacks, 
We train separate classifiers under \textsc{Attr-Privacy-Strict} and \textsc{Attr-Privacy-Relaxed} setups, ensuring alignment between the training objective and evaluation condition. Each classifier is trained in a 1-vs-all classification task for a given attribute and the goal is to decide whether a speaker attribute is present or not.


Transparent attacks expose severe leakage of speaker attributes. Pretrained embeddings combined with logistic regression achieve near-perfect separation of accent and gender on anonymized speech under \textsc{Attr-Privacy-Relaxed}. A deeper investigation into the interpretation of randomness in these results, as well as the behavior observed under \textsc{Attr-Privacy-Relaxed}, is left for future work.

These results reveal that the existing anonymization system may not be effective in protecting speaker traits. 
We also tested whether fine-tuning embedding models could strengthen attacker performance. In practice, we argue that fine-tuning with limited data samples could degrade the embedding geometry, which can lead to unstable evaluation outcomes. 
This degradation does not indicate improved privacy. In fact, since transparent attacks with pretrained embeddings already succeed, the attributes remain separable, and the fine-tuned models simply fail to exploit this due to distorted embeddings. 


\vspace{-2mm}
\section{Conclusion}
Our results show that user-privacy leakage is far more severe than what EER suggests. While EER remains stable, low-FPR analysis demonstrates that informed adversaries can achieve near-perfect precision for subsets of users. For attribute privacy, anonymization fails to protect sensitive traits such as accent and gender, as transparent attacks achieve near-perfect separation even on anonymized speech. These findings highlight the need to evaluate anonymization under worst-case and low-FPR conditions, and to develop methods that explicitly protect demographic attributes. In future work, we plan a theoretical analysis of low-FPR privacy bounds and to establish \texttt{VoxGuard} as a benchmark for comparing anonymization methods.

\newpage

\section{Acknowledgements}
This work was supported by the Office of the Director of National Intelligence (ODNI), Intelligence Advanced Research Projects Activity (IARPA), via the ARTS Program under contract D2023\-2308110001 and the HIATUS Program under contract \#2022\-22072200006. The views and conclusions contained herein are those of the authors and should not be interpreted as necessarily representing the official policies, either expressed or implied, of ODNI, IARPA, or the U.S. Government. The U.S. Government is authorized to reproduce and distribute reprints for governmental purposes notwithstanding any copyright annotation therein.

\bibliographystyle{IEEEbib}
\bibliography{strings,refs}

\begin{thebibliography}{10}

\bibitem{gdpr2016}
European {U}nion {EU},
\newblock ``{G}eneral {D}ata {P}rotection {R}egulation,'' \url{https://gdpr-info.eu/}, 2016.

\bibitem{zhang2023voicepm}
Shaohu Zhang, Zhouyu Li, and Anupam Das,
\newblock ``Voicepm: A robust privacy measurement on voice anonymity,''
\newblock in {\em Proceedings of the 16th ACM Conference on Security and Privacy in Wireless and Mobile Networks}, 2023, pp. 215--226.

\bibitem{haq2009speaker}
Sanaul Haq,
\newblock ``Speaker-dependent audio-visual emotion recognition,''
\newblock {\em personal. ee. surrey. ac. uk}, 2009.

\bibitem{tayebi2024addressing}
Soroosh Tayebi~Arasteh, Tom{\'a}s Arias-Vergara, Paula~Andrea P{\'e}rez-Toro, Tobias Weise, Kai Packh{\"a}user, Maria Schuster, Elmar Noeth, Andreas Maier, and Seung~Hee Yang,
\newblock ``Addressing challenges in speaker anonymization to maintain utility while ensuring privacy of pathological speech,''
\newblock {\em Communications Medicine}, vol. 4, no. 1, pp. 182, 2024.

\bibitem{ahangaran2025obfuscation}
Meysam Ahangaran, Nauman Dawalatabad, Cody Karjadi, James Glass, Rhoda Au, and Vijaya~B Kolachalama,
\newblock ``Obfuscation via pitch-shifting for balancing privacy and diagnostic utility in voice-based cognitive assessment,''
\newblock {\em Alzheimer's \& Dementia}, vol. 21, no. 3, pp. e70032, 2025.

\bibitem{tomashenko2024voiceprivacy}
Natalia Tomashenko, Xiaoxiao Miao, Pierre Champion, Sarina Meyer, Xin Wang, Emmanuel Vincent, Michele Panariello, Nicholas Evans, Junichi Yamagishi, and Massimiliano Todisco,
\newblock ``The voiceprivacy 2024 challenge evaluation plan,''
\newblock {\em arXiv preprint arXiv:2404.02677}, 2024.

\bibitem{nautsch2020privacy}
Andreas Nautsch, Jose Patino, Natalia Tomashenko, Junichi Yamagishi, Paul-Gauthier No{\'e}, Jean-Fran{\c{c}}ois Bonastre, Massimiliano Todisco, and Nicholas Evans,
\newblock ``The privacy zebra: Zero evidence biometric recognition assessment,''
\newblock 2020.

\bibitem{carlini2022membership}
Nicholas Carlini, Steve Chien, Milad Nasr, Shuang Song, Andreas Terzis, and Florian Tramer,
\newblock ``Membership inference attacks from first principles,''
\newblock in {\em 2022 IEEE symposium on security and privacy (SP)}. IEEE, 2022, pp. 1897--1914.

\bibitem{shokri2017membership}
Reza Shokri, Marco Stronati, Congzheng Song, and Vitaly Shmatikov,
\newblock ``Membership inference attacks against machine learning models,''
\newblock in {\em 2017 IEEE symposium on security and privacy (SP)}. IEEE, 2017, pp. 3--18.

\bibitem{dwork2006differential}
Cynthia Dwork,
\newblock ``Differential privacy,''
\newblock in {\em International colloquium on automata, languages, and programming}. Springer, 2006, pp. 1--12.

\bibitem{feng2022user}
Tiantian Feng, Raghuveer Peri, and Shrikanth Narayanan,
\newblock ``User-level differential privacy against attribute inference attack of speech emotion recognition in federated learning,''
\newblock {\em arXiv preprint arXiv:2204.02500}, 2022.

\bibitem{wang2024asynchronous}
Rui Wang, Liping Chen, Kong~Aik Lee, and Zhen-Hua Ling,
\newblock ``Asynchronous voice anonymization using adversarial perturbation on speaker embedding,''
\newblock {\em arXiv preprint arXiv:2406.08200}, 2024.

\bibitem{shamsabadi2022differentially}
Ali~Shahin Shamsabadi, Brij Mohan~Lal Srivastava, Aur{\'e}lien Bellet, Nathalie Vauquier, Emmanuel Vincent, Mohamed Maouche, Marc Tommasi, and Nicolas Papernot,
\newblock ``Differentially private speaker anonymization,''
\newblock {\em arXiv preprint arXiv:2202.11823}, 2022.

\bibitem{noe2021adversarial}
Paul-Gauthier No{\'e}, Mohammad Mohammadamini, Driss Matrouf, Titouan Parcollet, Andreas Nautsch, and Jean-Fran{\c{c}}ois Bonastre,
\newblock ``Adversarial disentanglement of speaker representation for attribute-driven privacy preservation,''
\newblock 2021.

\bibitem{vauquier2025legally}
Nathalie Vauquier, Brij Mohan~Lal Srivastava, Seyed~Ahmad Hosseini, and Emmanuel Vincent,
\newblock ``Legally validated evaluation framework for voice anonymization,''
\newblock in {\em Proc. Interspeech 2025}, 2025, pp. 3229--3233.

\bibitem{thudi2022bounding}
Anvith Thudi, Ilia Shumailov, Franziska Boenisch, and Nicolas Papernot,
\newblock ``Bounding membership inference,''
\newblock {\em arXiv preprint arXiv:2202.12232}, 2022.

\bibitem{casanova2024xtts}
Edresson Casanova, Kelly Davis, Eren G{\"o}lge, G{\"o}rkem G{\"o}knar, Iulian Gulea, Logan Hart, Aya Aljafari, Joshua Meyer, Reuben Morais, Samuel Olayemi, et~al.,
\newblock ``Xtts: a massively multilingual zero-shot text-to-speech model,''
\newblock {\em arXiv preprint arXiv:2406.04904}, 2024.

\bibitem{koizumi2023libritts}
Yuma Koizumi, Heiga Zen, Shigeki Karita, Yifan Ding, Kohei Yatabe, Nobuyuki Morioka, Michiel Bacchiani, et~al.,
\newblock ``Libritts-r: A restored multi-speaker text-to-speech corpus,''
\newblock {\em arXiv preprint arXiv:2305.18802}, 2023.

\bibitem{feng2025vox}
Tiantian Feng et~al.,
\newblock ``Vox-profile: A speech foundation model benchmark for characterizing diverse speaker and speech traits,''
\newblock {\em arXiv preprint arXiv:2505.14648}, 2025.

\bibitem{yamagishi2019vctk}
Junichi Yamagishi, Christophe Veaux, and Kirsten MacDonald,
\newblock ``Cstr vctk corpus: English multi-speaker corpus for cstr voice cloning toolkit,'' Version 0.92, [sound], 2019.

\bibitem{diwan2025scaling}
Anuj Diwan, Zhisheng Zheng, David Harwath, and Eunsol Choi,
\newblock ``Scaling rich style-prompted text-to-speech datasets,''
\newblock {\em arXiv preprint arXiv:2503.04713}, 2025.

\bibitem{champion2023anonymizing}
Pierre Champion,
\newblock ``Anonymizing speech: Evaluating and designing speaker anonymization techniques,''
\newblock {\em arXiv preprint arXiv:2308.04455}, 2023.

\bibitem{pantel1998spamcop}
Patrick Pantel, Dekang Lin, et~al.,
\newblock ``Spamcop: A spam classification \& organization program,''
\newblock in {\em Proceedings of AAAI-98 workshop on learning for text categorization}. Citeseer, 1998, pp. 95--98.

\bibitem{kantchelian2015better}
Alex Kantchelian, Michael~Carl Tschantz, Sadia Afroz, Brad Miller, Vaishaal Shankar, Rekha Bachwani, Anthony~D Joseph, and J~Doug Tygar,
\newblock ``Better malware ground truth: Techniques for weighting anti-virus vendor labels,''
\newblock in {\em Proceedings of the 8th ACM Workshop on Artificial Intelligence and Security}, 2015, pp. 45--56.

\bibitem{ho2017detecting}
Grant Ho, Aashish Sharma, Mobin Javed, Vern Paxson, and David Wagner,
\newblock ``Detecting credential spearphishing in enterprise settings,''
\newblock in {\em 26th USENIX security symposium (USENIX security 17)}, 2017, pp. 469--485.

\bibitem{desplanques2020ecapa}
Brecht Desplanques, Jenthe Thienpondt, and Kris Demuynck,
\newblock ``Ecapa-tdnn: Emphasized channel attention, propagation and aggregation in tdnn based speaker verification,''
\newblock {\em arXiv preprint arXiv:2005.07143}, 2020.

\end{thebibliography}

\end{document}